\documentclass[12pt]{article}
\begin{document}
\baselineskip .3in
\begin{titlepage}
\begin{center}
{\large{\bf On The Prediction of  The Statistical Model}}\vskip
.2in
  S. N. Banerjee, R. Ghosh and A. Bhattacharya $^{\ddag}$,
\end{center}

\vskip .1in
\begin{center}
Department of Physics, Jadavpur University \\
Calcutta 700032, India.\\
\end{center}
\vskip .1in
\begin{center}
{\bf Abstract}
\end{center}
 \ \vskip .3in

 The number of quark flavours $n_{f}$, has been derived to be
 equal to 6 in agreement with the experimental observations. The
 colour factor ratio $\frac{C_{A}}{C_{F}}$ has also been computed
 and is found to be in exact agreement with the corresponding QCD
 prediction. Further, the $e^{+}e^{-}$ pair annihilation cross
 section into hadrons at very high energy is found to be a
 function only of the fractal dimension of the hadron.

  \vskip .2in
 $^{\ddag}$ e-mail: pampa@phys.jdvu.ac.in
\end{titlepage}

{\bf I. Introduction}:

\vskip .2in

The statistical model, since its very inception three decades ago,
has undergone various applications with success $[1,2,3]$. The
model has a deep rooted concept of a relation between quark matter
and the space embedded in it and has explained and predicted a
number of interesting and bizzare features $[4,5,6]$. The
assignment of the (asymptotic) fractal dimension $D=\frac{9}{2}$
to a hadron has been the crucial factor in analysing the various
properties of hadron. Most recently $[7]$, we have studied the
colours of quarks as new degrees of freedom and the number of
colours equal to 3 has been derived.Further, the fractional nature
of the electric charges of quarks has also been suggested. In the
current investigation, we have derived the number of quark
flavours $n_{f}$,the colour factor ratio $\frac{T_{F}}{C_{F}}$.
The $e^{+}e^{-}$ pair annihilation cross section into hadrons at
high energy has been studied.

 \vskip.2in

 {\bf II. Theory}:

\vskip.2in

To deduce the number of quark flavours $n_{f}$ and to estimate the
magnitude of the well known QCD colour factor ratio
$\frac{T_{F}}{C_{F}}$,we are to study the production of hadrons in
the final state initiated by a gluon producing gluons and quarks
in the intermediate stages corresponding to the processes 1 and 2
respectively.In the intermediate stages,each energetic gluon or
quark creates a jet of hadrons in the final state.However,gluon
jets are broader than those in the case of quarks as the gluons
can emit bremsstrahlung themselves and this is more pronounced
because of greater colour charge.With the increase of the energy
of the quarks in the quark jet,the number of quark flavours grows
to its maximum value $n_{f}$ and there would be a close similarity
between quark and gluon jets.

\vskip.1in
   As usual,we denote$[8]$ the relative strengths of the splitting
   probabilities with $C_{A}$ and $T_{F}$ for gluon splitting into
   two gluons and two quarks respectively.If the final state is
   not differentiated with respect to its flavour content,an
   additional factor $n_{f}$ has to be taken into account as the
   gluon can split into $n_{f}$ quark flavours.Both the gluon and
   quark jets in processes 1 and 2 respectively produce
   hadrons,each having the fractal dimension D as suggested by the
   model.As fractal dimension is a geometrical property of metric
   transformations,it remains invariant under such a
   transformation$[9]$.Consquently,the two spaces corresponding to
   the quark and gluon jets at $Q^{2} \rightarrow \alpha $ where
   $Q^{2}$ is the square of the characteristic energy scale,would
   behave like equivalent metricspaces.Hence the spaces would be
   topologically equivalent i.e. homeomorphic.Unlike isotopy,the
   two diagrams corresponding to the aforesaid two processes would
   thus become homeomorphic as it would depend on the diagrams
   themselves and not on their disposition in space.Hence we come
   across the vertex $V_{1}$ representing the three gluon coupling
   corresponding to the process 1 and the vertex $V_{2}$
   representing the coupling of the gluon with two quarks
   corresponding to the process 2.Therefore,the vertex $V_{2}$ at
   high energy limit would be mapped into the vertex $V_{1}$.Hence
   the vertex factor $n_{f}T_{F}$ in the limit $Q^{2} \rightarrow \alpha
   $at $V_{2}$ would approach the vertex factor $C_{A}$ at
   $V_{1}$.Therefor,we have
\vskip .1in
\begin{equation}
   Lt._{Q^{2}\rightarrow\alpha}     n_{f}T_{F}\rightarrow C_{A}
\end{equation}
 \vskip .1in

 It is relevant to assert that we have also derived in the
 framework of the model$[7]$,the magnitude of the colour factor
 $C_{A}=3$,from the fractal dimension of hadron,without any
 reference to the experimental findings.Using $C_{A}=3$ directly
 in (1),we have $n_{f}=6$ since $T_{F}=\frac{1}{2}$.Further,from
 (1),we get
\vskip .1in
\begin{equation}
 Lt._{Q^{2}\rightarrow\alpha}      n_{f}\frac{T_{F}}{C_{F}} = \frac{C_{A}}{C_{F}} \frac{9}{4}
\end{equation}
 \vskip .1in
 where we have used our previously derived value of
 $\frac{C_{A}}{C_{F}}$ from fractal dimension$[5]$.With our estimate
 of $n_{f} = 6$ as an input in (2),we get
\vskip .1in
\begin{equation}
\frac{T_{F}}{C_{F}} = \frac{3}{8}
\end{equation}
 \vskip .1in
 in exact agreement with the corresponding QCD prediction.It is
 worth mentioning that the aforesaid value of $\frac{T_{F}}{C_{F}} = \frac{3}{8}$
in (3) has been obtained from the fractal properties of hadron
suggested by model.

 \vskip .1in
It is well-known that the cross section $\sigma_{e^{+}e^{-}}
\rightarrow $ hadrons coincides with the cross section
$\sigma_{e^{+}e^{-}} \rightarrow q\overrightarrow{q}$ at very high
energies from quark hadron duality and that $\sigma$ for
$e^{+}e^{-}$ annihilation into hadronic final states has the
form$[8]$
 \vskip .1in
\begin{equation}
\sigma = f(\alpha_{s}C_{F}, \frac{C_{A}}{C_{F}},
n_{f}\frac{T_{F}}{C_{F}})
\end{equation}
 \vskip .1in

This is,it is a function of the quark gluon coupling constant
$\alpha_{s}$ in addition to the other colour factors and $n_{f}$.
 \vskip .1in
The cross section for the process $e^{+}e^{-}\rightarrow$ hadron
is dominated by processes leading to the production of
$q\overrightarrow{q}$ pairs followed by a strong interaction
fragmentation process which converts the high energy
$q\overrightarrow{q}$ pair into jets of hadrons.As $Q^{2}
\rightarrow \alpha $,we have $\alpha_{s}\rightarrow 0$ and
$n_{f}\frac{T_{F}}{C_{F}}$ becomes equal to $\frac{C_{A}}{C_{F}} =
\frac{D}{2}= \frac{9}{4} $.Therefore,we arrive at
\vskip .1in
\begin{equation}
 Lt._{Q^{2}\rightarrow\alpha}  \sigma = f(\frac{D}{2})
\end{equation}
 \vskip .1in

 Thus the quark hadron duality in conjunction with the model
 suggests that the asymptotic $e^{+}e^{-}$ pair annihilation cross
 section becomes a function of the fractal dimension of hadron.

 \vskip.2in

 {\bf III. Summary:}

\vskip.2in

 one of the most interesting results of the current investigation
 is the prediction of the number of quark flavours $n_{f}$ equal
 to 6 without any reference to the experimental findings.The
 colour factor ratio $\frac{T_{F}}{C_{F}}$ derived is in exact
 agreement with the corresponding QCD prediction and the
 asymptotic $e^{+}e^{-}$ pair annihilation cross section is found
 to depend only on the fractal dimension of hadron.

\newpage

\vskip 0.2in {\bf References}

\vskip .2in

\noindent [1]. S. N. Banerjee et al., Had. J. {\bf 4}(1981) 203;
{\bf (E) 5},(1982)2157;{\bf 6} (1983)440,760; {\bf 11}(1989)243;
{\bf 12}(1989)179;{\bf 13}(1990)750.

 \noindent [2]. S. N. Banerjee et al., Int. J. Mod. Phys. {\bf A
 16}, No. 3 (2001)201.

 \noindent [3]. S. N. Banerjee et al., Int. J. Mod. Phys. {\bf A
 17}, No. 3 (2002)4939.

\noindent [4]. S. N. Banerjee and S. Banerjee, Phys. Lett. {\bf B
644 }(2007)245.

\noindent [5]. S. Mukhererjee and S. Banerjee, Mod. Phys. Lett.
{\bf A 24} No. 7(2009).

\noindent [6].  S. Mukhererjee and S. Banerjee,
arXiv:hep-ph/1011.4024.

 \noindent [7].  S. Mukhererjee and S.
 Banerjee,arXiv:hep-ph/1104.1534.

\noindent [8]. G. Dissertori et al., Quantum Chromodynamics,
Clarendon Press, Oxford, (2003)23.

\noindent [9]. M. F Barnsley, Superfractals, Cambridge University
Press, Cambridge (2006),27.

\end{document}